# Understanding Learner-LLM Chatbot Interactions and the Impact of Prompting Guidelines

Cansu Koyuturk[1,4], Emily Theophilou[2], Sabrina Patania[1], Gregor Donabauer[3], Andrea Martinenghi[1], Chiara Antico[1], Alessia Telari[1], Alessia Testa[1], Sathya Buršić[1], Franca Garzotto[1,4], Davinia Hernandez-Leo[2], Udo Kruschwitz[3], Davide Taibi[5], Simona Amenta[1], Martin Ruskov[6], and Dimitri Ognibene[1,7](✉)

[1] Università degli Studi di Milano Bicocca, Milan, Italy
dimitri.ognibene@unimib.com
[2] Universitat Pompeu Fabra, Barcelona, Spain
[3] University of Regensburg, Regensburg, Germany
[4] Politecnico di Milano, Milan, Italy
[5] Istituto per le Tecnologie Didattiche (ITD-CNR), Palermo, Italy
[6] Università degli Studi di Milano, Milan, Italy
[7] Institute of Science and Technology of Cognition (ISTC-CNR), Rome, Italy

**Abstract.** Large Language Models (LLMs) have transformed human-computer interaction by enabling natural language-based communication with AI-powered chatbots. These models are designed to be intuitive and user-friendly, allowing users to articulate requests with minimal effort. However, despite their accessibility, studies reveal that users often struggle with effective prompting, resulting in inefficient responses. Existing research has highlighted both the limitations of LLMs in interpreting vague or poorly structured prompts and the difficulties users face in crafting precise queries. This study investigates learner-AI interactions through an educational experiment in which participants receive structured guidance on effective prompting. We introduce and compare three types of prompting guidelines: a task-specific framework developed through a structured methodology and two baseline approaches. To assess user behavior and prompting efficacy, we analyze a dataset of 642 interactions from 107 users. Using *Von NeuMidas*, an extended pragmatic annotation schema for LLM interaction analysis, we categorize common prompting errors and identify recurring behavioral patterns. We then evaluate the impact of different guidelines by examining changes in user behavior, adherence to prompting strategies, and the overall quality of AI-generated responses. Our findings provide a deeper understanding of how users engage with LLMs and the role of structured prompting guidance in enhancing AI-assisted communication. By comparing different instructional frameworks, we offer insights into more effective approaches for improving user competency in AI interactions, with implications for AI literacy, chatbot usability, and the design of more responsive AI systems.

## 1 Introduction

Large Language Models (LLMs) have transformed the landscape of conversational agents by enabling natural language-based communication with AI-powered chatbots. With the advent of systems such as ChatGPT, they can craft relevant and coherent responses to inquiries, mirroring human language patterns [1]. This paradigm shift has democratized access to advanced conversational technologies, allowing individuals from diverse backgrounds to utilize AI for a wide range of applications, from information retrieval [2], text summarization [3] to creative writing [4], without the necessity of having any formal training in how to use these systems.

Studies have shown that individuals face difficulties in formulating efficient prompts, determining efficacy, and understanding the technical aspects of interactions with LLMs [5]. Particularly, inexperienced users may struggle with crafting prompts that are both clear and sufficiently detailed [6]. Furthermore, responses generated by LLMs can sometimes be unclear or difficult to interpret [7]. These challenges can be further compounded when LLMs generate excessively verbose or ambiguous responses in reaction to vague or poorly structured inputs. Given these challenges, research suggests that structured approaches to prompting can help bridge the gap between user intent and AI comprehension by providing users with better guidance, interfaces, and interaction methods [8]. Providing users with targeted prompting guidelines can enhance their ability to communicate effectively with LLMs, leading to improved response accuracy and usability [9].

The growing prevalence of instruction-based LLMs [10, 11] raises critical questions about their role in education, specifically, how these systems can be integrated into teaching, how learners interact with AI, and what skills educators must develop to use them effectively. Previous research has explored various ways to prompt ChatGPT to function as an educational chatbot [12], examined high school student perceptions that shape interactions with such systems [13], and evaluated different versions of ChatGPT in assessing these interactions [14]. This work seeks to explore the potential of a similar adoption in a hybrid human-AI learning environment, with the overall aim of investigating the possibilities and limitations of LLM-enriched educational contexts to provide an efficient learning experience. This research investigates the effectiveness of prompting strategies for LLMs by adapting the secondary education classroom study [13] to an online context, expanding both the prompting techniques and the generalizability of prior findings beyond young learners. Participants are trained to use LLMs and perform prompting exercises, allowing us to explore variants of skills and knowledge that enable individuals to confidently and effectively use AI and reflect on their experiences; learn to use AI; learn to communicate and collaborate with AI; and understand limitations and problems of AI. In addition, we evaluate the quality of interactions by systematically categorizing user and AI responses using Von NeuMidas, an extended pragmatic annotation schema that captures both the semantic and pragmatic dimensions of human-LLM interactions [15]. Von NeuMidas is particularly useful for analyzing complex, free-form commands issued by users, as well as assessing the LLM's ability to interpret and respond to such inputs effectively.

## 2 Background

Several strategies for improving users' prompting skills have been proposed in the literature, drawing on various theoretical frameworks and practical insights. These strategies aim to guide learners in refining their interaction techniques during human-AI interactions. One prominent approach involves defining clear roles for both the user and the AI [16, 17]. In this framework, the user provides explicit instructions regarding how the AI should behave in the conversation, such as adopting a particular persona or tone. This strategy, based on the Persona pattern, is based on the idea that AI interactions can be more effective when the AI is guided to embody specific characteristics, such as those of a teacher, expert, or conversational partner. It is especially beneficial for educational prompting training, as it helps users learn to steer AI behaviour intentionally, leading to more relevant, structured, and meaningful interactions. Additionally, the use of flipped conversation patterns requires the LLM to ask questions rather than generate output [16]. This strategy flips the interaction flow, so the LLM asks the user questions to achieve a desired goal. As a result, the interaction becomes more focused and efficient, enabling the LLM to tap into knowledge that the user might not initially possess and achieve more accurate and timely outcomes. The 'Flipped Interaction' pattern is particularly beneficial for educational prompting training because its reliance on AI-driven questioning promotes active knowledge discovery, a key element of effective pedagogical practices.

Another strategy emphasizes structuring the conversation with clear, step-by-step instructions [12, 18]. This method helps users define the sequence of actions the AI should take during interactions, emphasizing personalization and natural conversational flow. Additionally, the prompt should account for the user's cultural background and knowledge level, allowing the AI's responses to adapt accordingly. Moreover, other works have emphasized the importance of simplicity in communication by utilizing multiple short sentences [17]. This approach encourages the specification of key subtopics of interest and the clear listing of actions that the chatbot must perform, ensuring that the interaction remains focused and effective for the user. Collectively, these practices can foster more effective and meaningful AI-assisted learning experiences.

Although various studies have suggested different methods for effective prompting, there is a noticeable gap in research focused on specific educational strategies for educating users towards improving their prompting skills. Additionally, few studies have explored which specific prompting techniques are most effective in enhancing user competency. To address the gap in research on effective educational strategies for improving prompting skills, frameworks like scaffolding and authentic learning offer valuable solutions. Scaffolding involves providing structured support that gradually decreases as learners gain proficiency [19], potentially allowing them to develop more effective prompting techniques. This structured guidance may also enhance engagement [20], ensuring that learners are not only exposed to AI tools but are also supported in learning how to interact with them in a meaningful way. Another relevant perspective is authentic learning, which emphasizes hands-on engagement with real-world tasks [21]. Rather than passively receiving information about AI, learners actively experiment with prompting strategies, iterating on their approaches based on the responses they receive. Engaging students in hands-on experiences with AI tools can potentially allow them to develop a deeper understanding of AI's capabilities, limitations, and ethical implications through direct experience [13, 22]. Furthermore, an iterative learning process, where learners repeatedly engage with AI, test different strategies, and refine their approach, while receiving immediate feedback on the effects their prompts produce, may support deeper understanding and long-term skill development [23]. By combining scaffolding with authentic learning, there is potential to create an approach that not only teaches learners how to prompt effectively but also encourages critical engagement with AI systems. In this study, we aim to design an authentic learning approach supported by scaffolded tasks, enabling participants to develop their prompting skills through the identified prompting strategies.

## 3 Methods

### 3.1 Design and Participants

This study adhered to the Helsinki Declaration and received ethical approval from the University of Milan-Bicocca's minimal risk research committee (Ethical Protocol no. RM-2023-715). All participants provided written informed consent before participation and were debriefed after data collection (10/2023–02/2024). A total of 107 participants (aged 18–32 years, $M = 25.00$, $SD = 3.00$; 62% female) residing in the UK or the USA were recruited through the Prolific platform [24] to take part in the experiment. Participants who rated their prior use of ChatGPT on a 5-point Likert scale, with limited experience defined as selecting "1" (never) or "2" (once or twice), were included in the study. All collected data was anonymized to protect participants' privacy. To collect the participants' prompts, we developed a custom interface using the ChatGPT API [12]. Built with TypeScript, React, NextJS, and SQLite, the platform logs user interactions, including inputs, ChatGPT responses, and metadata like timestamps, in a SQL database for analysis.

The study began with an introductory video that addressed general AI literacy, which included a comparative analysis of AI versus human intelligence and discussed the capabilities and limitations of generative AI (GenAI) (see **Fig. 1** for an overview of the complete procedure). Following the training, all participants engaged in three consecutive interactions with ChatGPT focused on a non-trivial educational activity. Specifically, they were instructed to prompt ChatGPT to act as a personal teacher, explaining the potential threats and underlying mechanisms of social media in a coherent and natural conversational manner. After this initial phase, participants were randomly assigned to one of three groups, each introduced to different prompting strategies aimed at enhancing interactions with ChatGPT (PS1 n=20, PS2 n=57, PS3 n=30, see **Table 1**). During the subsequent training sessions, participants received training on their assigned prompting guidelines, which included detailed descriptions and examples (see 3.2 Training on Prompting Strategies). Following the training, participants once again interacted with ChatGPT in three consecutive sessions on the same educational activity. A total of 642 interactions took place, during which 4536 messages were exchanged between the conversational agent and the participants.

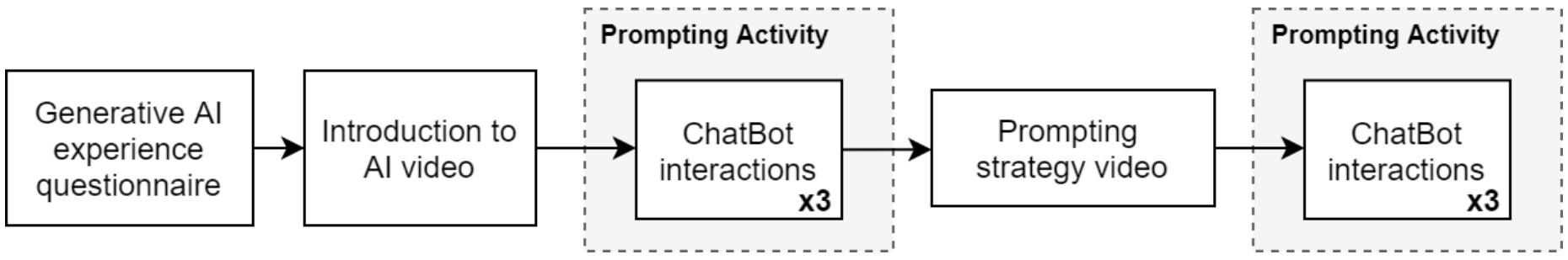

**Fig. 1.** An overview of the procedure of the study, previously published in [14].

### 3.2 Training on Prompting Strategies

The three prompting strategies were selected to represent varying levels of task specificity and generation cost (see **Table 1**): PS1 was developed ad hoc based on domain knowledge and analysis of interaction success [12, 13], sometimes on other forms of GenAI [17]. PS2 was drawn from existing literature on prompt pattern codification [16]. Lastly, PS3 was generated directly by ChatGPT, reflecting a low-effort, AI-derived task-specific alternative.

**Table 1.** The three proposed prompting strategies.

| PS1. Authors contributed | PS2. Derived From [16] | PS3. Suggested by ChatGPT |
|---|---|---|
| 1.1. Define conversational roles and how they should be played (e.g., "Act like an expert with friendly behavior") | 2.1. Persona Pattern (e.g., "Act as an exam grader") | 3.1. The Tone (e.g., "Behave as my fun and engaging teacher") |
| 1.2. Describe the conversation process providing a sequence of steps (e.g., "Ask a question, wait for my answer, then provide feedback") | 2.2. Flipped Conversation Pattern (e.g., "Ask me questions to define the assignment") | 3.2. Clarity and Specificity (e.g., "Explain one aspect of the topic") |
| 1.3. Prefer multiple simple sentences (e.g., "Use short, clear statements") | | 3.3. Use One-Step Instruction (e.g., "Explain this part first, then move on") |
| 1.4. Enumerate particular subtopics of interest (e.g., "Focus on topics A, B, and C") | | 3.4. Interactivity (e.g., "Ask me a question and wait for my reply") |
| 1.5. Specify all the actions that the chatbot must perform for a successful interaction (e.g., "List the steps to solve the problem and confirm each one") | | |

We begin with the first prompting strategy (PS1), task-specific guidelines, which defines key features of the desired ChatGPT behavior through guidelines focused on roles, processes, and ways to communicate. The LLM should be given clarity about both the user's and AI's roles (PS1.1), including how each role should be enacted. The interaction process should be outlined with defined steps (PS1.2) and a shared objective (PS1.5). If a more elaborate conversation is expected, prompts should also specify how the AI should participate in turn-taking to avoid conflicts. To prevent ambiguity as prompts grow more complex, the strategy recommends keeping language simple (PS1.4), using short sentences when possible (PS1.3), and guiding the AI toward particularly relevant subtopics.

We devised a second set of guidelines for a prompting strategy (PS2) based on [16]. From the full list of 16 prompt patterns proposed, we identified two as particularly relevant to our goals: the Persona pattern (PS2.1), which prompts the model to elicit a set of attributes associated with a specified persona; and the Flipped Interaction pattern (PS2.2), which shifts the typical user-AI dynamic by having the AI ask targeted questions to achieve a defined goal.

The third prompting strategy (PS3) includes four key elements. The tone (3.1) encourages the use of polite and engaging language that can lead to more engaging and informative interactions. The second guideline (PS3.2) emphasizes giving clear, specific, and concise instructions about the interaction's goal and topic, which can provide a more structured format of the conversation that is easier to adopt for the AI tool. The third guideline (PS3.3) recommends breaking down instructions into one step at a time, helping the AI process and respond sequentially. Finally, interactivity (PS3.4) involves setting up a back-and-forth question–answer format to create more engaging, natural conversations.

Noteworthy, some of the guidelines within these strategies overlap to varying degrees. For instance, defining roles in PS1.1 closely resembles the Persona pattern in PS2.1, while the use of simple sentences in PS1.3 aligns with the emphasis on clarity in PS3.2. However, these similarities are only partial, as each guideline functions within the broader framework of its respective strategy. Consequently, we did not find it meaningful to merge all strategies into a single comprehensive approach.

### 3.3 Annotation Process

To evaluate the effectiveness of human-LLM interactions and the application of prompting guidelines, a subset of dialogues was manually annotated. This process aimed to determine whether participants followed the task of the study, successfully applied the prompting strategies they were taught, and which interactions and guidelines yielded the best results. The evaluation was conducted in three key steps.

**Interaction Description Features.** The preliminary evaluation focused on two aspects: prompt success (whether participants attempted to achieve the objective of the study's task) and task success (whether the interaction achieved the intended learning activity). The annotation criteria for both prompt and task success were defined in a group session, ensuring consistency in evaluation. Each of these two steps included four specific features example of a prompt success feature (e.g. "The user provides a clear task definition to the conversational agent within the initial or second input."), and example of a task success feature (e.g.: "The conversation consistently stays on topic, discussing potential threats and mechanisms of social media "). Since this process was found to be clear and straightforward, annotations were conducted by four annotators, with each annotator reviewing a single interaction in these two steps.

During the evaluation of prompt success, the initial and second inputs of participants were both examined, as some participants began conversations with a greeting rather than directly stating the task to the conversational agent. Each annotator identified ten well-structured and ten poorly structured examples for each prompting strategy. A prompt was classified as a successful prompt attempt if it contained two or more predefined features from the group discussion. In the second step, interactions that met this criterion were further assessed for task success, which was determined by evaluating the entire conversation. The interaction was deemed successful if it satisfied three or more of the predefined features.

**Prompting Guidelines.** The annotation process evaluated inter-annotator agreement across three guideline types: Task-Specific, Pattern, and GPT-Generated Guidelines. A total of twelve prompts were manually annotated by four annotators, each of whom was familiar with the annotation framework, and each prompt was reviewed by two annotators to ensure reliability. Using Cohen's κ as a measure of agreement, the results show clear differences in annotation consistency. Task-Specific Guidelines achieved the highest agreement (κ=0.853), followed by Pattern Guidelines (κ=0.741), and GPT-Generated Guidelines had the lowest agreement (κ=0.703) (see **Table 2**). This suggests that more structured, manually designed guidelines improve annotation consistency, while AI-generated guidelines may require refinement to ensure reliability.

The variability in agreement can be attributed to several factors. Task-Specific Guidelines provided clear, well-defined rules, reducing ambiguity and leading to near-perfect agreement in some cases. Pattern Guidelines, while systematic, required contextual interpretation, which may have introduced subjectivity, leading to moderate agreement. GPT-Generated Guidelines, on the other hand, exhibited the highest level of disagreement, with a minimum Cohen's κ as low as 0.429. This suggests that AI-generated instructions may lack clarity or specificity, making them more difficult to apply consistently. A final agreed annotation, determined by a third expert resolving disagreements, served as the gold standard for evaluating LLM-based annotations for both prompting guidelines and Von NeuMidas annotation processes.

**Table 2.** Comparison of Agreement Across Guidelines.

| Metric | Task-specific | Patterns | ChatGPT-Generated |
|---|---|---|---|
| Feature-wise κ Average | 0.841 | 0.698 | 0.600 |
| Overall Guideline κ | 0.853 | 0.741 | 0.703 |
| Min κ in Guideline | 0.636 | 0.478 | 0.429 |
| Max κ in Guideline | 1.000 | 0.831 | 0.831 |

**Von NeuMidas Annotation Process for Dialogues.** As the final step, we manually annotated 12 interactions, four from each prompting guideline, using the newly defined Von NeuMidas pragmatics annotation scheme [15], specifically designed to integrate established conversational approaches with the unique characteristics of instruction-tuned LLMs. Four annotators participated, with each interaction evaluated by two annotators. The overall agreement across all speech acts with MIDAS [25] was high (κ=0.921), indicating strong inter-annotator reliability. However, agreement levels varied significantly across different speech act categories, with some achieving near-perfect agreement and others showing little to no reliability.

Speech acts such as “positive-answer” (κ=0.97), “thanks” (κ=0.95), and “opening” (κ=0.94) had the highest agreement, suggesting that these acts were well-defined and consistently recognized by annotators. Similarly, “task-command” (κ=0.75) and “statement-non-opinion” (κ=0.91) showed strong reliability, likely because they involve structured, direct communication. Conversely, categories such as “general-opinion” (κ=−0.004), “negative-answer” (κ=−0.0012), and “yes-no-question” (κ=−0.0024) had near-zero or even negative agreement, indicating discrepancies in interpretation and high annotator uncertainty in classifying these speech acts. Distinguishing between MIDAS labels can be challenging because overlapping linguistic cues and ambiguous phrasing often obscure the differences among factual, opinion, and yes-no questions. Moreover, polite forms of commands, commonly used even in interactions with chatbots equipped with advanced natural language abilities, further complicate matters by blending indirect request markers with directive language. This blending makes it difficult to decide whether these utterances should be classified as valid task commands or dismissed as invalid commands.

In contrast, the Von Neumann components of the annotation had low consistency. This is in part because they are applied only to initiative speech acts, i.e. questions and commands, so their number was limited. Moreover, although annotators most often used the "output command" label, they sometimes disagreed on whether to label the chatbot’s general or task-specific suggestions as commands, especially when these suggestions didn’t clearly lead to an observable response during the dialogue. Moreover, the difficulty in distinguishing between output and input commands, with inverted roles of executor and addressee, strongly affected their annotation consistency. This led to a revision of the annotation schema and a careful creation of the gold standard to enable the few-shot learning process execution and evaluation. When considering only the initial part of the conversation, the von Neuman commands’ annotation got an overall Cohen’s κ agreement of 0.79.

### 3.4 Automation of Annotations with LLM

We automated the annotation process of the human–LLM interactions, we adopted the few shot learning methodology for a similar condition outlined in [14], with 2 positive and 2 negative samples from the human annotated gold standard and employed the current release of GPT-4o (Feb 2025). To accelerate query processing, we deployed 48 parallel threads and implemented a strategy of repeated attempts with incremental delays approach effectively managed out-of-format responses and the API’s per-minute request limit. Under this configuration, tasks including prompt annotation, task success evaluation, guideline adoption, and generation of notations were completed in approximately eight hours. Although repeated sampling was not performed, due to time and cost constraints, a similar but more targeted approach may be beneficial for users seeking a more precise analysis on specific aspects of the interactions.

# 4 Analyses and Results

## 4.1 Evaluation of LLM annotations

For system accuracy evaluation, we employed a leave-one-out methodology. Each sample from the expert-annotated gold standard was classified five times using a few-shot training paradigm. Specifically, for each classification, two positive and two negative samples were randomly selected from the gold standard, excluding the sample currently under evaluation. Accuracy varied across different annotation components; for instance, MIDAS speech acts exhibited the highest error rate, reflecting their inherent complexity displayed also during the annotation process. Overall, the average accuracy across features was 0.73, with individual accuracies ranging from 0.51 to 0.94. These results are consistent with those reported in [14]. A comprehensive investigation into the causes and variability of annotation complexity across dimensions is beyond the scope of this paper.

## 4.2 Results of Pre- and Post-Training Phases for Prompt and Task Success

Since the training was organized into two distinct phases, we examined changes in user behavior by calculating the average performance across the three sessions in each phase, with these averages reflecting both adherence to the task (prompt success) and the overall quality of the human-LLM conversations (task success).

Jamovi (V 2.4.14.0) was used to analyze the data and repeated measures ANOVA were conducted separately for 2 aspects. Results indicated a significant main effect of training on performance on both prompt success ($F(1, 93) = 8.447$, $p = .005$, $\eta^2 = 0.083$) and task success ($F(1, 104) = 10.237$, $p = .002$, $\eta^2 = .090$). Participants demonstrated a statistically significant increase in crafting successful prompts and the overall quality of the human-LLM conversations was improved following the training. However, the interaction between training and prompting guidelines was not statistically significant for either prompt success ($F(2, 93) = 0.198$, $p = .821$, $\eta^2 = 0.004$) or task success $F(2, 104) = 0.128$, $p = .880$, $\eta^2 = .002$). This indicates that the effect of training on successfully creating prompts and the overall quality of the human-LLM conversations did not vary significantly across the 3 different prompting guidelines.

## 4.3 Results for the Application of Prompting Guidelines

We evaluated the effectiveness of the three prompting guidelines in detail by totalling the frequency with which users incorporated the specified features into their prompts. After the training, each group exhibited improvements that aligned with the objectives of the corresponding guidelines. Specifically, for the ChatGPT-generated features, there was an increase in the use of one-step instructions and specificity of topics meaning that participants demonstrated a shift toward clearer, more organized requests which enabled the AI to process and respond more effectively. For the pattern prompting guidelines, participants exhibited a substantial rise in the use of Flipped Patterns, which suggests that users were able to engage in more controlled and interactive dialogues aimed at creating an interaction to facilitate learning.

The changes in the use of each feature of the task-specific guideline can be seen in **Fig. 2** (A). After training, participants demonstrated a shift away from concise topic statements and less reliance on simple sentence structures, indicating that they were able to articulate their requests more clearly and directly. They adopted a more detailed description of the task as pointed out by the increase in the number of goals presented in the prompt. Meanwhile, the increase in role play and role form/context requests demonstrates that users increasingly specified the perspective or style in which the AI should generate its responses which likely contributed to generating content that was more relevant and aligned with user expectations. The correlation between feature counts before and after training and task success offers an understanding of which features perform more effectively and whether users have enhanced their implementation strategies (see **Fig. 2** (B)). Overall, the training appears to have encouraged users to formulate clearer, more purposeful interactions with AI. Also, it is interesting to note that not effective features, that were discouraged in the guidelines had negative correlation with success and decreased their counts after training.

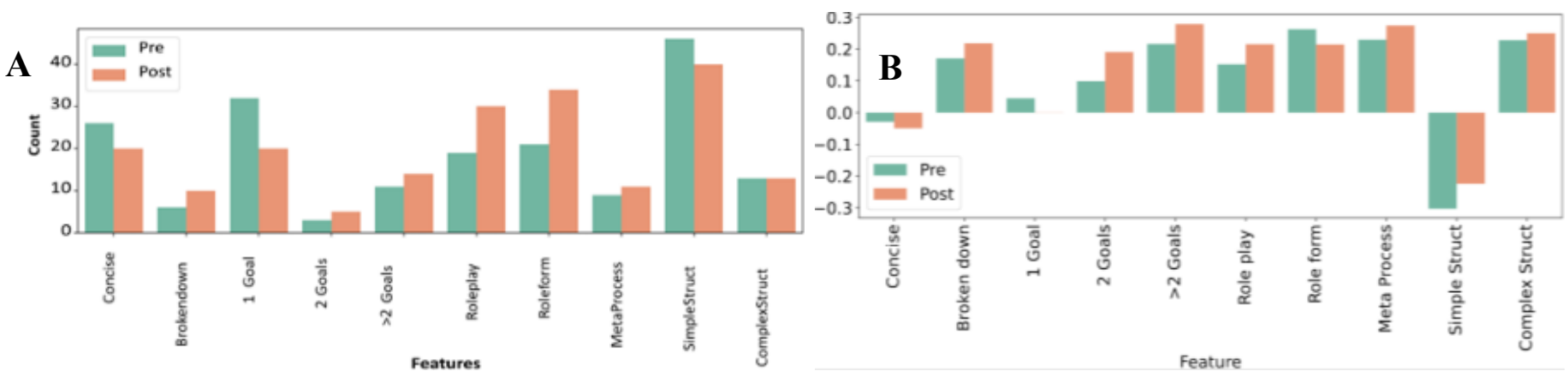

**Fig. 2.** Results for Task-specific guidelines: Feature count (A): Difference between before and after the related prompting strategy training; Correlation between features and task success (B).

### 4.4 Results for the Interaction Quality with Von NeuMidas

While the data confirm a typical user–assistant dynamic, users often request information, and the assistant primarily provides answers, there are clear asymmetries in how social conventions, questions, and answers are distributed (See **Fig. 3**) [26, 27]. Humans engage in minimal social convention (~13% after a factual question, which are likely offers of assistance), whereas the assistant uses social conventions more frequently, responding with over 30% to such acts or even to answers or commands. Even if LLM chatbot's task is to lead the conversation [28], users still drive most interactions: assistant answers prompt new user questions or commands (>70%), whereas user answers lead to about 40% opinions and further answers from the assistant. Likewise, users respond to opinion questions with other questions 65% of the time and offer direct answers only 19%, while the assistant provides answers to opinion questions 57% of the time. Assistant factual questions elicit follow-up questions more than half the time (~51%), whereas user factual questions receive a 54% direct response and about 33% follow-up questions or commands. Additionally, the self-transition of statements indicates monologue-like sequences, underscoring that both users and the assistant can engage in extended discourse. To delve deeper into these conversational patterns and asymmetries, future research might adopt formal methods such as random walks over these transition probabilities [29, 30].

Von NeuMidas effectively captures shifts in user behavior post-training, offering a task-independent lens for analysing user interactions. As shown in **Fig. 4** (A), the MIDAS feature counts indicate a rise in user-issued directions and a slight decrease in responses (answers, statements, opinions), suggesting a more direct and less exploratory interaction style. There is also an increase in social conventions, potentially reflecting greater comfort or adherence to the ChatGPT-generated guidelines advocating polite interactions. Meanwhile, **Fig. 4** (B) highlights a general uptick in all Von NeuMidas command types, particularly output commands, implying that users are more inclined to let the agent steer the conversation. The surge in state commands similarly points to users specifying how the interaction should proceed or how the agent should respond, underscoring a more proactive and directive approach post-training. These findings showcase the utility of Von NeuMidas annotations in analyzing and understanding human–LLM interactions; further task-agnostic insights could be gleaned from deeper analysis but are left for future work due to space constraints.

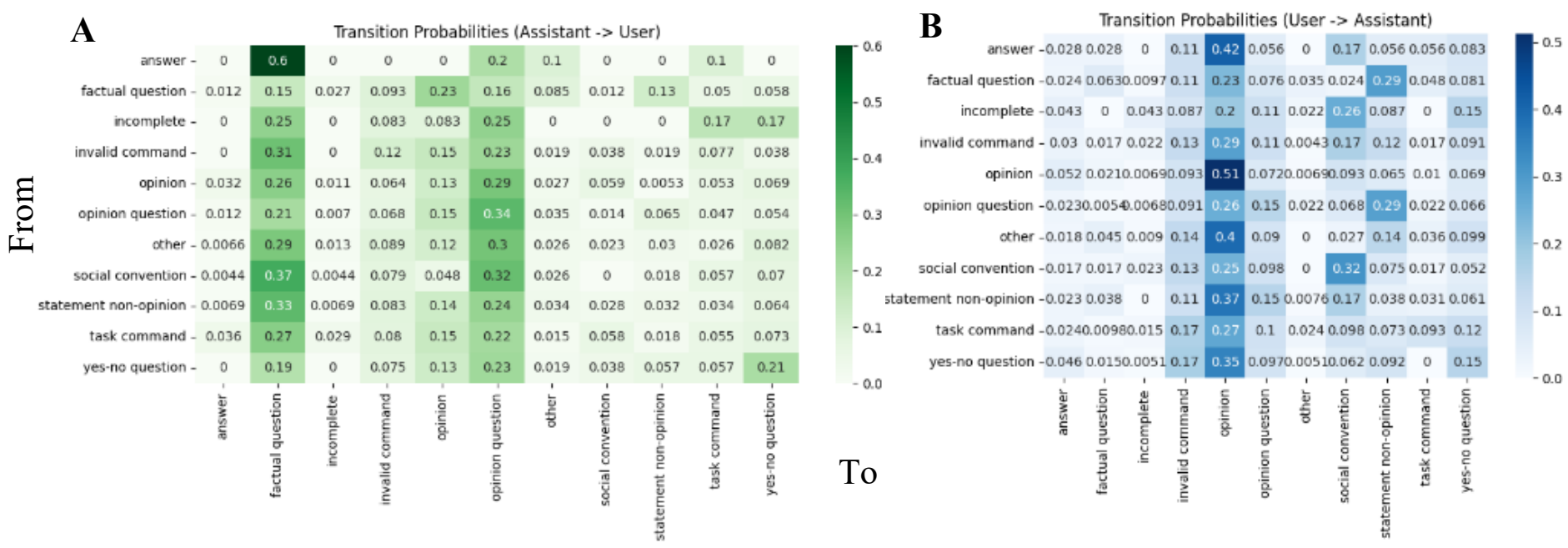

**Fig. 3.** Probabilities MIDAS speech acts: from assistant to user (A), from user to assistant (B).

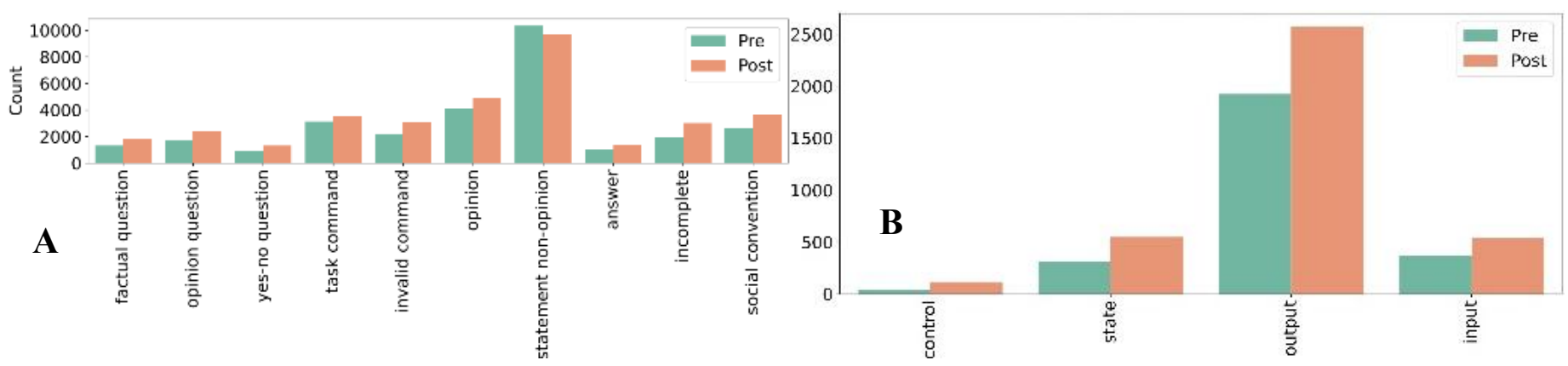

**Fig. 4.** User: Midas (A) and Von Neumann (B) features before and after training.

## 5 Discussion and Conclusion

Human-LLM interactions have become increasingly important in a variety of contexts. In simpler tasks, such as image generation, the process is more constrained, and defining a method is more straightforward [17]. However, due to their incremental nature, both the design of the initial prompt and the ongoing process of improvement and clarification are important. Therefore, we aimed to enhance AI literacy and introduced three distinct prompting guidelines that teach novice users how to effectively leverage and communicate with AI during complex tasks. Through hands-on practice, participants were able to iteratively refine their approaches and receive immediate feedback, thereby directly observing the impact of their adjustments in an authentic learning environment.

Our results showed that training enabled users to create more organized and precise requests, which allowed the conversational agent to process and respond more effectively. These improvements in prompts resulted in clearer, more purposeful interactions, which also provides further evidence of the significance of prompt engineering training. Moreover, our findings point to the value of Von NeuMidas annotations for generalizing beyond task-specific performance. The schema effectively captured shifts in user behavior after training, particularly in how participants issued more directive and structured prompts. These patterns reflect a more intentional and goal-oriented interaction style, supporting the use of Von NeuMidas for uncovering task-agnostic peculiarities and insights about human–LLM interactions. These results underscore the high potential of the collected public dataset to support further research on prompting practices and interaction dynamics.

However, we did not find statistically significant differences between the three prompting strategies. This outcome may be attributed to the prompting strategies partially overlapping in their features, which could have reduced between-group differences. Additionally, the unequal sample sizes across groups may have reduced statistical power, limiting our ability to detect more subtle effects. These discrepancies in group sizes stemmed from technical issues and varying participant dropout rates, likely also influenced by the overall length of the experiment. Notably, our sample included individuals with limited prior experience with AI. As such, future studies should explore how expert users, who may already possess refined prompting skills, respond to similar forms of prompting instruction. It is also important to consider that our training focused on prompting techniques and did not address other essential components of GenAI literacy, such as considerations of algorithmic bias, misinformation, and overreliance [31]. A comprehensive training approach that integrates these elements is necessary to ensure a well-rounded and ethically informed engagement with GenAI systems.

Our findings suggest that future AI-based educational tools should incorporate integrated tutorials or tooltips that guide users on how to formulate clear and specific prompts for different learning tasks for clearer and purposeful interactions. Interestingly, the observed training effects also suggest that prompting guidelines can be devised effectively across various task contexts with limited effort.

**Acknowledgments.** This work was supported by Erasmus+ KA220-HED Cooperation Partnerships in Higher Education (IDEAL project, no. 2024-1-IT02-KA220-HED-000251425) and the Volkswagen Foundation (COURAGE project, no. 95567) as well as by PID2020-112584RB-C33, PID2023-146692OB-C33, CEX2021-001195-M funded by MICIU/AEI/10.13039/501100011033 and SGR 00930. DHL (Serra Húnter) also acknowledges the support by ICREA Academia.